\documentclass[conference]{iacrtrans}  % arXiv version
\usepackage[utf8]{inputenc}

\usepackage{hyperref}
\usepackage{cleveref}
\usepackage{todonotes}
\author{
  Elie Bursztein\inst{1} \and
  Michael Gruber\inst{5} \and
  Karel Král\inst{2} \and
  Jean-Michel Picod\inst{2} \and
  Matthias Probst\inst{4} \and
  Georg Sigl\inst{3,4}
}
\institute{
  Google, Sunnyvale, USA, \email{scaaml@google.com}
  \and
  Google, Zurich, Switzerland, \email{scaaml@google.com}
  \and
  School of Computation, Information and Technology, Technical University of Munich, Munich, Germany, \email{matthias.probst@tum.de}
  \and
  Technical University of Munich
  TUM School of Computation, Information and Technology, \email{georg.sigl@aisec.fraunhofer.de}
  \and
  Fraunhofer Institute for Applied and Integrated Security (AISEC), Munich, Germany, \email{michael.gruber@aisec.fraunhofer.de}
}
\authorrunning{E.Bursztein \and M.Gruber \and K.Král \and J-M Picod \and M.Probst \and G.Sigl} % J-M without dot is the preferred by J-M.

\title{Profiling Resilient to Change in Probe Position}

\begin{document}

\maketitle

%%%% 5. KEYWORDS %%%%

%%%% 6. ABSTRACT %%%%
\begin{abstract}
    Side Channel Analysis (SCA) relaxes the black-box assumption of conventional cryptanalysis by incorporating physical measurements acquired during cryptographic operations.
    Electro-magnetic (EM) emissions of a chip during computations often provide a very valuable source of side channel leakage.
    During the evaluation of a chip for electro-magnetic side channel emissions one needs to position an electro-magnetic probe in an advantageous position relative to the chip.
    Previous literature focused on hot-spot finding and to a lower extend repositioning.
    Trace augmentations have been considered to aid portability of profiling using one physical device and attacking another device.
    This paper focuses on training a single neural network using traces from multiple EM probe positions to detect leakage from a larger area over the attacked device.
    We provide dual evaluation of EM traces – from two completely independent labs – profiling on data from one lab and attacking traces from the other lab.
\end{abstract}

%%% Custom commands:
%% Lab position -- anonymous submission
%\newcommand{\labMun}{Lab~A}
%\newcommand{\labZrh}{Lab~B}  % The desk
%\newcommand{\labZrhBasement}{Lab~$\beta$}
%% Final version -- not anonymous
\newcommand{\labMun}{Munich~lab}
\newcommand{\labZrh}{Zurich~desk}
\newcommand{\labZrhBasement}{Zurich~basement}
%% Dataset names -- anonymous submission
%\newcommand{\datasetSwAesMun}{SW-AES-A}
%\newcommand{\datasetSwAesZrh}{SW-AES-B}
%% Dataset names -- not submission
\newcommand{\datasetSwAesMun}{SW-AES-Mun}
\newcommand{\datasetSwAesZrh}{SW-AES-Zrh}
\newcommand{\datasetHwAesTrain}{HW-AES-T}
\newcommand{\datasetHwAesHoldout}{HW-AES-H}
\newcommand{\datasetHwAesHoldoutLarger}{HW-AES-H2}
\newcommand{\datasetSecondLookSTM}{SL-STM}
\newcommand{\datasetSecondLookXMEGA}{SL-XM}

% Used in 02_background.tex
% Consider using: https://de.overleaf.com/learn/latex/Glossaries
\newcommand{\gls}[1]{{#1}}

%%%% 7. PAPER CONTENT %%%%
\section{Introduction}\label{sec:introduction}

In many situations it is advantageous to use the electromagnetic emanations as a source of side channel information.
These can be much more precisely targeted at a specific part of the device under test and thus greatly reduce the noise coming from activity unrelated to the computation of interest.
However they come with their own set of challenges.
Profiled EM attacks are generally not considered very portable due to the fact that one would need to find a very similar relative position of the chip and EM probe.
The question arises whether the research is reproducible.
Even when a research group publishes a dataset they would also need to publish a precise relative position.
This could in theory be done on a de-processed chip using a camera and manual repositioning but very few publications considered this setting.
The state of the art largely hurts not only reproducibility but also continuous testing during development.
In case more data is needed later but the measurement setup has been modified for other campaigns in the meantime it is not an efficient strategy to discard all previous data.

\emph{Non-profiled attacks} such as CPA~\cite{CHES:BriClaOli04} attempt to directly extract information out of a given device under test without any profiling.
This makes them immune to any EM probe re-positioning issues as there is no profiling chip and thus no discrepancy between the profiling position and the attack position or between the profiling and attacked chip.
Much has been argued about the strengths and weaknesses of profiled vs non-profiled analysis and whether the assumption of having a profiling device is too strong or not.
We believe that both approaches are valuable.
However to compare these fairly in the context of EM leakage one must consider issues of probe repositioning and cross-device portability.

\emph{Previous work} has produced results for repositioning, some negative and some positive results for portability, and some trace adaptation techniques.
Often these results were conducted with a large-diameter EM probes and not very secure implementations.
See \Cref{sec:related_work} for more details.

\emph{Our contributions} are the following:

We profile on traces from multiple EM probe positions to better adapt to traces from the attack device where the relative position of the EM probe and chip might not be fully known.
This method has been suggested before by~\cite{EPRINT:GMDDSR20} as an interesting future research.

% What sets us apart
Thanks to cross-institution collaboration we are the first to show a profiled attack with profiling data coming from a completely different lab -- a completely different equipment located in two different countries and without ever visiting each other's lab.

We show that these results are obtainable on more convoluted implementations (hardware AES on STM32F415).
Due to not having the same equipment this dataset has been captured on two setups by the same team but in different rooms and using different equipment (only the EM probe was shared).
To battle very low leakage we develop a new hybrid technique of using a neural network as a signal amplifier for CPA.
This technique is of independent interest.

All of our experiments are conducted on off-the-shelf ChipWhisperer target boards (STM32F3 and STM32F4) without any de-capping.
Moreover, these boards have their decoupling capacitors removed to ease power attacks at the cost of harder EM attacks.
\section{Background}\label{sec:background}

A prominent subclass of \gls{SCA} is power analysis (\gls{PA}).
Differential power analysis (\gls{DPA}) exploits a large number of power traces and a statistical distinguisher to recover secret information by correlating measured leakage with hypothetical models~\cite{Kocher1999}.
To mitigate the combinatorial growth of hypotheses, \gls{DPA} attacks typically employ a divide-and-conquer strategy in which the secret is decomposed into small subcomponents that are attacked independently and subsequently recombined.
%Targeting compact intermediate values keeps the hypothesis space tractable.

An effective targeted subcomponent has a sufficiently small domain to provide a managable hypothesis size and depends on both secret and public data so that competing hypotheses yield distinguishable leakage predictions.
For varying public inputs, hypotheses are instantiated for the selected intermediate value and mapped to hypothetical power consumption using a leakage model, such as the Hamming weight or Hamming distance (\gls{HW} or \gls{HD}) model.
%While Kocher~\etal~\cite{Kocher1999} originally proposed the difference-of-means as a distinguisher, Pearson’s correlation coefficient was later introduced by Brier~\etal~\cite{Brier2004} and has since become very popular due to its improved statistical efficiency and robustness.

For \gls{DPA} or \gls{CPA}, an adversary applies $n$ distinct public inputs $D$ to the device under test (\gls{DUT}) and records the corresponding power traces while the cryptographic function $f(D, s_c)$ is evaluated under the secret parameter~$s_c$.
The hypothesis space of candidate secrets is denoted by $\mathcal{S}=\{s_0,s_1,\dots,s_{|\mathcal{S}|-1}\}$, where $s_c\in\mathcal{S}$ represents the true secret.
For each candidate $s_j\in\mathcal{S}$ and each public input $D$, a hypothetical intermediate value is computed by simulating the relevant algorithmic component, i.e., by evaluating a reduced function $f'(D,s_j)$ following the divide-and-conquer approach.
These intermediate values are mapped to predicted leakage using a leakage model, yielding a hypothesis matrix $\mathbf{H}\in\mathbb{Z}^{|\mathcal{S}|\times n}$. %, where each row $\mathbf{h}_{j,\cdot}$ corresponds to the predicted leakage of candidate $s_j$ across all measurements.
Each hypothesis is statistically compared to the recorded traces using a distinguisher.
In case of \gls{CPA}, this is Pearson's correlation coefficient \cite{Brier2004}.
The candidate whose predicted leakage exhibits the strongest statistically significant agreement with the measurements is identified as the correct secret $s_c$, assuming the presence of exploitable data-dependent leakage.

\paragraph{Machine Learning Side Channel Analysis}

While \gls{CPA} relies on an explicit power model to compare hypotheses with measured traces, \gls{MLSCA} replaces this model by training an \gls{ML} algorithm to infer the processed data directly from observed traces.
As a training-based approach, \gls{MLSCA} belongs to the class of profiled \gls{SCA}, in which a separate training device is used to learn a model that is subsequently applied to classify traces from the target device.
Ideally, the training device is identical to the target and executes the same algorithm, while the secret data is fully controlled by the adversary to enable supervised training \cite{Hospodar2011, Picek2017}.

Let $\mathbf{T}\in\mathbb{R}^{n\times m}$ denote the matrix of measured power traces, where $n$ is the number of measurements and $m$ the number of sampled time points, and let $\mathcal{S}=\{s_0,\dots,s_{|\mathcal{S}|-1}\}$ be the set of possible secrets.
In a profiled setting, a training device is used to collect a labeled dataset
\[
\mathcal{D}_{\mathrm{train}} = \{(\mathbf{t}_i, s_i)\}_{i=1}^{n_{\mathrm{train}}},
\]
where each trace $\mathbf{t}_i\in\mathbb{R}^m$ is associated with a known secret label $s_i\in\mathcal{S}$.

A machine-learning model $g_\theta(\cdot)$ with parameters $\theta$ is trained to approximate the posterior distribution
\[
p(s \mid \mathbf{t}) \approx g_\theta(\mathbf{t}),
\]
by minimizing an empirical loss function over $\mathcal{D}_{\mathrm{train}}$.
After training, the model is applied to traces collected from the target device, yielding for each trace $\mathbf{t}_i$ a vector of class scores or posterior probabilities $\hat{p}(s_j \mid \mathbf{t}_i)$ for all $s_j\in\mathcal{S}$.

For a set of $n_{\mathrm{test}}$ target traces, evidence is aggregated across measurements, for example by summing log-likelihoods,
\[
\Lambda_j = \sum_{i=1}^{n_{\mathrm{test}}} \log \hat{p}(s_j \mid \mathbf{t}_i),
\]
and the recovered secret is identified as
\[
s_c = \arg\max_{s_j\in\mathcal{S}} \Lambda_j,
\]
provided that the learned model captures exploitable data-dependent leakage shared between the training and target devices.
\section{Threat Model}\label{sec:threat_model}

We are following the standard profiled side channel assumptions \cite{CHES:ChaRaoRoh02} where the attacker has full access to a clone of the targeted device.
Furthermore, we assume that the attacker has a set of tools to acquire EM traces from an untouched chip (no de-processing is allowed).
The EM probe diameters are $250\mu m$ and $500 \mu m$.
The positioning precision during the attack is limited to relative positioning of the EM probe and the chips packaging.

The only scenario considered by our work is white-box meaning the evaluator has full control over the profiling device and the inputs and outputs of the cryptographic primitive.
Having a model trained on the profiling device the evaluator attempts to capture similar traces from the targeted device \emph{while replicating the capture setups using different physical chip, oscilloscope, positioning equipment, and accessories of the same type} and use the trained model to recover the secrets.
\section{Related Work}\label{sec:related_work}

%%%%%%%%%%%%%%%%%%%%%%%%%%%%%%%%
% EM attacks beginning and SoK %
%%%%%%%%%%%%%%%%%%%%%%%%%%%%%%%%
%\cite{zunaidi2024systematic} Systematic Literature Review of EM-SCA Attacks on Encryption
%\cite{EPRINT:PPMWB21} {SoK}: Deep Learning-based Physical Side-channel Analysis
Electromagnetic emanations have been suggested as a potential side channel by~\cite{jean2000new}.
Initially unpublished work dates all the way back to~\cite{nacsim19825000}.
A rich body of research has been published since.
See literature review papers~\cite{EPRINT:PPMWB21,zunaidi2024systematic} for more references.
%\cite{specht2019high} High Resolution EM Side Channel Attacks with Multiple Measurement Probes
%%%%%%%%%%%%%%%%%%%%%%%%%%%%%%%%
% Probe positioning strategies %
%%%%%%%%%%%%%%%%%%%%%%%%%%%%%%%%
%\cite{jiang2021probe} A Probe Placement Method for Efficient Electromagnetic Attacks
Finding the leakage, optimizing the EM probe position, and automating this process is an active area of research, e.g., \cite{jiang2021probe} and \cite{mehta2025swarm}.

%%%%%%%%%%%%%%%%%%%%%%%%%%%%%%%%%%%
% power side channel cross device %
%%%%%%%%%%%%%%%%%%%%%%%%%%%%%%%%%%%
%\cite{EPRINT:DGDGRS19} X-{DeepSCA}: Cross-Device Deep Learning Side Channel Attack
%\cite{NDSS:BCHJPS20} Mind the Portability: {A} Warriors Guide through Realistic Profiled Side-channel Analysis

Portability of profiled side channel analysis has often been overlooked.
Even relatively recent works \cite{EPRINT:DGDGRS19,NDSS:BCHJPS20} suggest that portability of power-based side channel attacks does not come for free.
The landscape of portability of profiled EM-based SCA research can be roughly categorized into the following classes:

%%%%%%%%%%%%%%%%%%%%%%%
% Probe repositioning %
%%%%%%%%%%%%%%%%%%%%%%%
%\cite{EPRINT:RicWilMor19} Automated Probe Repositioning for On-Die {EM} Measurements
%\cite{EPRINT:GMDDSR20} Efficient Electromagnetic Side Channel Analysis by Probe Positioning using Multi-Layer Perceptron
\emph{Probe repositioning} has been first studied by \cite{EPRINT:RicWilMor19} to determine the relative position of an EM probe and device under test.
The authors trained a deep neural network that given a trace returns an estimate of relative position (regression of the $x,y$ coordinates).
They used on-die measurements of the STM32L0 target by the Langer ICR HH150-27 probe and estimated the repositioning error to be roughly $50\mu m$.
The repositioning error is roughly in the order of magnitude of the diameter and resolution of the probe given by the manufacturer: $150 \mu m$ and $100 \mu m$, respectively.
Unfortunately there is very little literature which would quantify the influence of these positioning errors on SCA.

Repositioning was formulated as a classification problem on a grid (predicting a discrete grid point rather than a continuous position in a rectangle) by \cite{EPRINT:GMDDSR20}.
They worked with an intact CW308T-XMega target using a $6 \times 6$ grid with a step of $2mm$ and about $1mm$ above the chip surface.
Measurements were taken using an H field probe with a coil diameter of $10mm$ (TBPS01 EMC Near-Field Probe).

Statistical properties of traces combined with various image matching methods have been used by~\cite{probenav} to speed up the repositioning process.

%%%%%%%%%%%%%%%%%%%%%%%
% Position resilience %
%%%%%%%%%%%%%%%%%%%%%%%
%\cite{navanesan2024ensuring} Ensuring cross-device portability of electromagnetic side-channel analysis for digital forensics
%Trace-to-Trace Translation for SCA \cite{genevey2021trace}
%Train or Adapt a Deeply Learned Profile \cite{LC:GenHeuGer21}
%On What to Learn: Train or Adapt a Deeply Learned Profile? \cite{EPRINT:GenGerHeu20}
\emph{Portability of profiled analysis}, namely cross-device and cross-model portability, was studied by~\cite{navanesan2024ensuring}.
For their smartphone experiments the authors used various iPhones and attempted to identify which of a given set of 10 activities is running (e.g., calendar, email, ...).
Due to differences present even in identical devices the authors had to use transfer learning techniques.
Their results with classifying which program is running on an nRF52-DK were similar.

Transfer between EM probe position and type (Langer near-field RF-B 0,3-3 and RF-K 7-4) can be successful on the same device as reported by \cite{genevey2021trace} using an AtMega8515 micro-controller running an ASCAD implementation of AES-128 (with a masking countermeasure).

%%%%%%%%%%%%%%%%%%%
% Adapting traces %
%%%%%%%%%%%%%%%%%%%

%\cite{danial2021x} Em-x-dl: efficient cross-device deep learning side-channel attack with noisy em signatures
%\cite{JCEng:MBTL13} Improving cross-device attacks using zero-mean unit-variance normalization
%\cite{TCHES:CZLG21} Cross-Device Profiled Side-Channel Attack with Unsupervised Domain Adaptation
%\cite{ninan2024second} A Second Look at the Portability of Deep Learning Side-Channel Attacks over EM Traces

\emph{Adapting traces} using various preprocessing techniques to reduce the differences between devices is a natural approach followed by various teams.
A combination of pre-processing techniques like PCA, LDA, and FFT with a training device selection algorithm has been used by \cite{danial2021x} to show a cross-device attack on an 8-bit ATXmega128D4.
\cite{JCEng:MBTL13} achieve significant effectiveness improvements of cross-device template attacks using zero mean and unit variance normalization while targeting 16-bit PIC24F micro-controllers.

An unsupervised fine-tuning phase was introduced by \cite{TCHES:CZLG21} for the pre-trained model on Atmel XMEGA MCUs and SAKURA-G evaluation boards.
The fine-tuning phase takes the original profiling traces and the unlabeled attack traces and uses the technique minimizing a distribution distance metric known from domain adaptation~\cite{gretton2012kernel}.

%\cite{EPRINT:CCCDDD19} Deep Learning to Evaluate Secure {RSA} Implementations
The high potential of deep learning attacks against secure implementations of RSA (EAL4+ certified arithmetic co-processor) has been identified by \cite{EPRINT:CCCDDD19}.
The authors also report that profiling portability (profiling on a device A and attacking a device B) works well in their setting.

The most comprehensive evaluation of pre-processing techniques so far and a large-scale dataset has been published by~\cite{ninan2024second} on both 8-bit and 32-bit targets (AVR XMEGA and STM32F3) with random delays.

%%%%%%%%%%%%%%%%%%
%%%%%% TODO %%%%%%
%%%%%%%%%%%%%%%%%%
\section{Methodology}\label{sec:methodology}

Our goal is to ensure portability as much as possible.
One of our datasets has been captured half in one lab and half by the another lab in a completely different geographic location.
The other dataset could not have been replicated due to the same equipment not being available at one of the labs.
We plan to publish our datasets to aid more experiments.
The focus is on datasets with traces coming from multiple known EM probe positions -- a $(x, y, z)$ grid or just $(x, y)$ grid when the distance from the device under test -- the $z$ coordinate -- is fixed.
\Cref{tab:dataset_overview} contains an overview of used datasets.
Side channel analysis is given in \Cref{sec:dl_results,sec:classical_attacks}.

\subsection{Dataset Collection}\label{sec:dataset_collection}

First we identify a source of leakage using an SNR~\cite{RSA:Mangard04} grid search and then confirm that both CPA~\cite{CHES:BriClaOli04} works and a neural network can be trained there.
Then we capture a dataset of EM measurements from each place of an $n \times n$ grid centered around the source of leakage.
We capture a grid dataset at \labMun{} and with a different setup at \labZrh{}.
The repositioning was done by eye (a photo) without any hot-spot finding at \labZrh{}.
In all measurements we use an unprocessed chip (no de-capping) and off-the-shelf ChipWhisperer targets without a decoupling capacitor.
The removal of a decoupling capacitor is usually done to improve power consumption leakage usually at the expense of EM~leakage ultimately making our analysis harder.
Positioning of the EM probe was done using a Langer ICS~105 table~\cite{LangerICS105}.
The ChipWhisperer target board has been fixed in place using either clamps or adhesive tape.
Neither solution allows for precise replication of an existing setup.

\labMun{} used PicoScope~6402 (\datasetSwAesMun{}) while \labZrh{} used PicoScope~6424E (\datasetSwAesZrh{}, \datasetHwAesTrain{}, \datasetHwAesHoldout{}, \datasetHwAesHoldoutLarger{}) since these models were available.
We do not expect the oscilloscope model difference to play a significant role in our measurements.

\begin{table}
    \centering
    \resizebox{\textwidth}{!}{%cheat with table size
    \begin{tabular}{ccccccccc}
        Name & Author & Grid & Step [mm] & Splits & Chip & EM Probe \\
        \hline
        \datasetSwAesMun{} & \labMun{} & $11 \times 11 \times 2$ & $0.5$ & 12.8/1.6/1.4 & STM32F3 & HH 250-75~\cite{LangerProbeHH250u75} \\
        \datasetSwAesZrh{} & \labZrh{} & $11 \times 11 \times 2$ & $0.5$ & 15.8/6.2/6.2 & STM32F3 & HH 250-75~\cite{LangerProbeHH250u75} \\
        \datasetSwAesZrh{} fine & \labZrh{} & $101 \times 101 \times 1$ & $0.05$ & 0/0/10.4 & STM32F3 & HH 250-75~\cite{LangerProbeHH250u75} \\
        \datasetSwAesZrh{} hotspot & \labZrh{} & $1 \times 1 \times 1$ & $0.0$ & 7.9/0.025/0.025 & STM32F3 & HH 250-75~\cite{LangerProbeHH250u75} \\
        \datasetHwAesTrain{} & \labZrhBasement{} & $5 \times 5$ & 0.25 & 11.5/1.1/3.2 & STM32F4 & Riscure~\cite{RiscureProbeHPelmagDS1203A} \\
        \datasetHwAesHoldout{} & \labZrh{} & $5 \times 5$ & 0.25 & 0/0/3.2 & STM32F4 & Riscure~\cite{RiscureProbeHPelmagDS1203A} \\
        \datasetHwAesHoldoutLarger{} & \labZrh{} & $11 \times 11$ & 0.25 & 0/0/89.8 & STM32F4 & Riscure~\cite{RiscureProbeHPelmagDS1203A} \\
        \datasetSecondLookSTM{} & \cite{ninan2024second} & $3 \times 3$ & $2.3$ & 1.215/0.135/0.15 & STM32F3 & RF 7-4~\cite{LangerProbeRF74NearField} \\
        \datasetSecondLookXMEGA{} & \cite{ninan2024second} & $3 \times 3$ & $2.3$ & 1.215/0.135/0.15 & XMEGA & RF 7-4~\cite{LangerProbeRF74NearField}
    \end{tabular}
    }%resizebox
    \caption{
        Dataset overview.
        For more information see \Cref{sec:dataset_collection}.
        The column Splits contains the number of traces in Train / Test / Holdout splits in millions of traces.
    }
    \label{tab:dataset_overview}
\end{table}

\subsubsection{Software AES}\label{sec:dataset_collection_swaes}

\datasetSwAesMun{} and \datasetSwAesZrh{} contain measurements of electromagnetic emanations caused by a custom unprotected software implementation of AES128 running on an STM32F303RCT7 target.
Half of the dataset has been captured in \labMun{} and another half in \labZrh{}.
We specify the whole chip since initially one of the labs unknowingly used a different STM32F3 chip and we confirm that there are differences between STM32F303RCT7 and STM32F303RBT6 as noted by~\cite{cwsupporteddevices}.
The measurements for \datasetSwAesMun{} are collected using the Langer HH 250-75~\cite{LangerProbeHH250u75} through a bias-tee~\cite{LangerBiasTee706} to a $30dB$ amplifier~\cite{LangerPreAmplifier306} and $1V$ for the channel sensitivity and PicoScope~6402 and 8 bits of ADC resolution.
The bias-tee~\cite{LangerBiasTee706} acts as a bandpass filter of 500kHz-6GHz.
Interestingly \datasetSwAesZrh{} needed $2V$ channel sensitivity to avoid overflows.
\datasetSwAesZrh{} has been captured using PicoScope~6424E with 12bit ADC resolution.
The distance between the probe and the chip is less than $0.2mm$ in the closer half of the $11 \times 11 \times 2$ grid.
The farther half of the dataset is $0.25mm$ farther from the original while the grid step was $0.5mm$ and thus covering an area of $25mm^2$.
The distance was confirmed by eye and almost crashing into the chip.
The chip clock was fed by ChipWhisperer and was set to 10MHz and the oscilloscope sampling rate is 625MHz.
A single trace consists of 7,000 points capturing the whole AES computation.
We use the input of the first round S-BOX as our target values.

Due to differences in our capture scripts the \datasetSwAesMun{} dataset the X-axis is increasing left to right and the Y-axis is increasing bottom to top.
In the \datasetSwAesZrh{} the X-axis is also increasing left to right but the Y-axis is increasing top to bottom.
Another notable difference is that \labMun{} represented their traces as integer values of ADC while \labZrh{} represented voltage values.
This results in different amplitudes but since the ADC to voltage scaling is linear this difference poses no issue.
We normalize the trace values to the interval $[-1, 1]$ based on maximum and minimum over all traces captured by a concrete lab.
This is consistent with standard deep learning practices (see for instance \cite{chollet2021deep}).
As usual with software implementations of AES we attack the first round S-BOX values.

\subsubsection{Hardware Accelerated AES}\label{sec:dataset_collection_hwaes}

The ChipWhisperer HW-AES firmware was run on two STM32F415RGT6 targets with hardware acceleration for AES.
A smaller $5 \times 5$ grid was captured to make our experiments more tractable.
We also opted for smaller step since we expected hardware accelerated leakage to be more localized.
Thus we captured with a grid step of $0.25 mm$ covering an area of $1mm^2$.
Split sizes and grid dimensions are shown in \Cref{tab:dataset_overview}.
Repositioning was done using a photo of the previous setup without any optical aid.
The whole dataset was captured by \labZrh{} (using PicoScope~6424E) since we could not capture the other part of the dataset in the other place due to a lack of availability of an equivalent EM~probe.
Two setups were used to maximize portability and the capture campaigns were run in two different rooms (the lab denoted as \labZrhBasement{} and a work desk denoted as \labZrh{}).
The only piece of equipment shared by these two setups was a single Riscure EM probe~\cite{RiscureProbeHPelmagDS1203A} with the $0.5mm$ coil tip.
The probe was chosen as it gave us better SNR (still very low).
We chose of this EM probe to improve the signal quality in these specific settings.

We captured a holdout split in \datasetHwAesTrain{} since hyper-tuning was necessary and also to be able to test our hybrid approach.
For the leakage model we use the last round Hamming distance as suggested by \cite{cwTutorialT2P2HWAES}.
At first we tried to either directly predict the last round Hamming distance as a classification problem or as a regression.
The training was unstable and we resorted to using a hybrid approach described in \Cref{sec:hwaes_hybrid_approach}.
We saw leakage practically everywhere in our \datasetHwAesHoldout{} but the number of traces was not enough for a full key recovery.
Thus we also captured an $11 \times 11$ grid (covering $2.5 \times 2.5 mm$) in \datasetHwAesHoldoutLarger{}.

\subsubsection{Publicly Available Datasets}\label{sec:dataset_collection_public}

We additionally use datasets published by~\cite{ninan2024second} to foster comparability to existing research.
Similarly to~\cite{ninan2024second} we directly target the output of the first round S-BOX.
To isolate from the influence of trace adaptations and preprocessing techniques we limit ourselves only to the raw traces.
The two datasets \datasetSecondLookXMEGA{}  (AVR XMEGA), resp., \datasetSecondLookSTM{}  (STM32F3) come from chips target~X1, resp., target~S1 where only the center position of the grid has been captured.
A full $3 \times 3$ grid has been captured over the targets X2 and S2.
Thus we are limited in our experiments.
We choose training on the full grid and evaluating accuracy over the center position.
Such experiments unfortunately provide no information regarding position resilience but the other choice would limit us to profiling traces from a single location.
Due to the type of EM~probe used (EMV RF 7-4 Near-Field Probe with resolution of roughly $5mm$~\cite{LangerProbeRF74NearField}) the grid step (smallest distance of two grid points) is about $2.3mm$ and the area from which the probe captures signal is significantly larger than in our experiments.

\subsection{Multiple Place Training}\label{sec:multiplace_training}

Profiling with traces from multiple locations has been proposed by~\cite{EPRINT:GMDDSR20}.
Using the analogy of image classification normalizing and centering images used to be common in the early research such as in the MNIST dataset~\cite{lecun1998gradient}.
However the amount of diverse data and larger deep learning models start performing better when normalization becomes more challenging with real world photography images.
We argue that instead of trying to come up with ways to make the traces appear to be taken from the same place we just let the model automatically adjust to the available data.

As described in \Cref{sec:dataset_collection} we first identify a source of leakage and capture a grid around it.
We use training and validation split from each grid position to quantify how much leakage our deep learning model can detect.
Effectively we are creating a heatmap of leakage.
We conduct another profiling phase training with all grid points where there was non-trivial leakage detected.
This trained model is then evaluated on each grid point from the attack target.
\Cref{sec:dl_results} provides ablation experiments -- changing this methodology to see which aspects are beneficial.
\section{Deep-Learning Results}\label{sec:dl_results}

We start by presenting our main result according to \Cref{sec:methodology}.
Ablation studies then follow investigating influence of our methodology choices on the our analysis.
One can compare some of these results with those in \Cref{sec:classical_attacks}.
For our experiments we build upon the model of \cite{TCHES:BIKMPZ24} to predict attack points in either classification or regression mode.
We have done little or no hyper-tuning.
The patch size is chosen to be a number which both divides the trace length and is roughly close to its square root.
Merge filters are set to zero since our traces are short compared to the original paper.
We use the Adafactor~\cite{shazeer2018adafactor} optimizer and follow the GPAM tutorial\footnote{\url{https://google.github.io/sedpack/tutorials/sca/gpam/}} recommendations.
Hyperparameters for all our ML experiments are listed in \Cref{tab:ml_hyperparameters}.
It is possible that a hyper-parameter search would yield better results.

\begin{table}[]
    \centering
    \begin{tabular}{lrrrr}
        Hyperparameter     & SWAES   & HWAES   & SL-STM & SL-XM   \\
        \hline
        Batch size         & 64      & 256     & 64     & 64      \\
        Steps per epoch    & 400     & 500     & 400    & 400     \\
        Epochs             & 150     & 150     & 900    & 900     \\
        Learning rate      & 0.0005  & 0.0001  & 0.0005 & 0.0005  \\
        Merge filter 1, 2  & 0       & 0       & 0      & 0       \\
        Trace length       & 7,000   & 10,000  & 5,000  & 5,000   \\
        Patch size         & 70      & 100     & 100    & 100
    \end{tabular}
    \caption{Hyperparameters of the GPAM model~\cite{TCHES:BIKMPZ24}.}
    \label{tab:ml_hyperparameters}
\end{table}

As usual in deep learning we normalize the trace values into the interval $[-1, 1]$ by linear scaling using the maximum and minimum over all traces regardless of grid position.
However due to different capture settings between \datasetSwAesMun{} (integer ADC values) and \datasetSwAesZrh{} (floating point voltage values) we need to scale by different factors.

\subsection{Portability and Position Resilience}\label{sec:portability_and_resilience}

In this section we present our main experiment showing both portability between different setups and chips and resilience to changes in the EM~probe position.

\begin{table}
    \centering
    \setlength{\tabcolsep}{0pt}  % No spacing between columns
    \def\figWidth{.24\linewidth}  % width of each figure
    \def\rotateOrigin{t}
    \begin{tabular}{cccc}
         \datasetSwAesMun{} close &
         \datasetSwAesMun{} far &
         \datasetSwAesZrh{} close &
         \datasetSwAesZrh{} far \\
         \includegraphics[width=\figWidth]{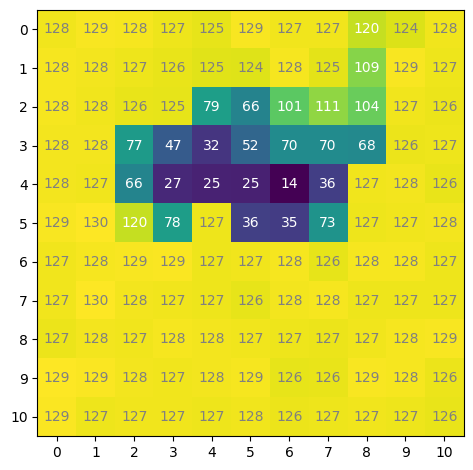} &
         \includegraphics[width=\figWidth]{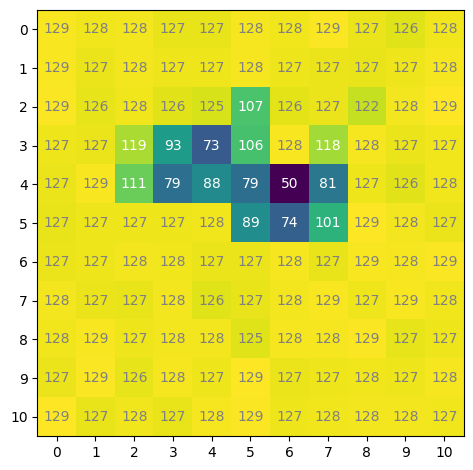} &
         \includegraphics[width=\figWidth]{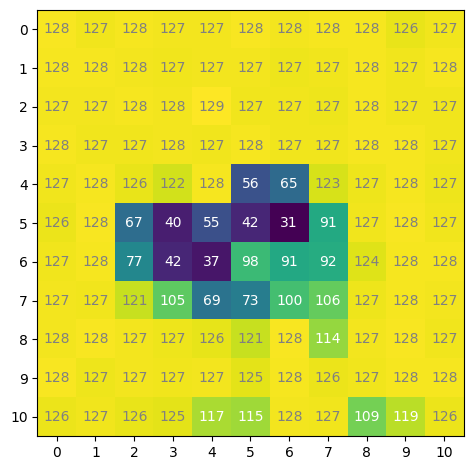} &
         \includegraphics[width=\figWidth]{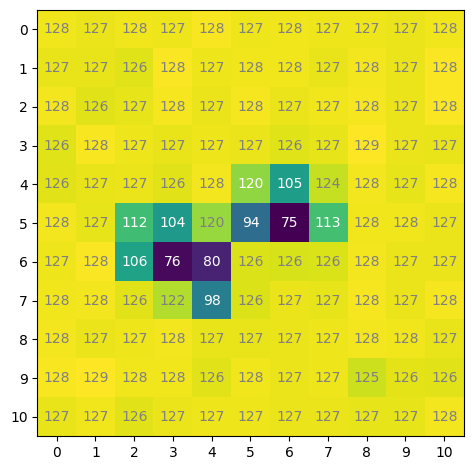} \\
    \end{tabular}
    \caption{
        Results of training from a single grid point on the software AES128.
        Values are averages of validation mean ranks during the second half of training targeting the index 11 of the S-BOX input in the first round.
    }
    \label{tab:sw_aes_grid_leakages}
\end{table}

For our \datasetSwAesMun{} and \datasetSwAesZrh{} datasets we choose to target the values of S-BOX input of the first round, i.e., the key xor plain-text.
No significant change in leakage detection or different levels of leakage was observed between different indices.
However doing multi-class learning for all 16 indices at once was roughly six to seven times slower.
Thus we restrict ourselves to the index 11 which was chosen at random.
\Cref{tab:sw_aes_grid_leakages} shows the leakages from a single grid point.
To combat possible over-fitting and to give more accurate results we present the averages of valuation results from the second half of the training.
Our selection parameter for a significant leakage is the mean rank value at most 120 (with $127.5$ being the random choice).
Any grid point with lower mean rank is selected for the position resilient training.
This selection strategy resulted in roughly $20\%$ of grid points from the closer and $12\%$ from the farther set to be used.

\begin{table}
    \centering
    \setlength{\tabcolsep}{0pt}  % No spacing between columns
    \def\figWidth{.24\linewidth}  % width of each figure
    \def\rotateOrigin{t}
    \begin{tabular}{cccc}
         \datasetSwAesMun{} close &
         \datasetSwAesMun{} far &
         \datasetSwAesZrh{} close &
         \datasetSwAesZrh{} far \\
         \includegraphics[width=\figWidth]{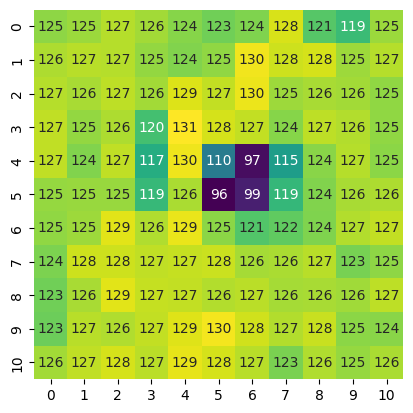} &
         \includegraphics[width=\figWidth]{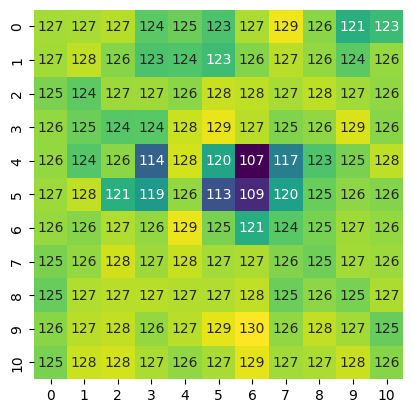} &
         \includegraphics[width=\figWidth]{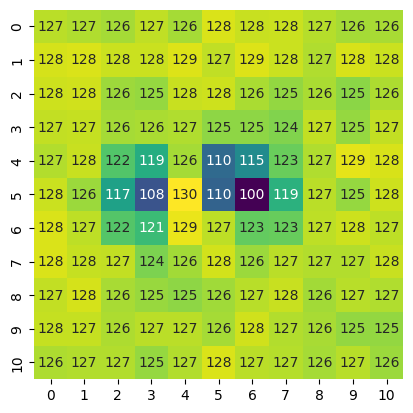} &
         \includegraphics[width=\figWidth]{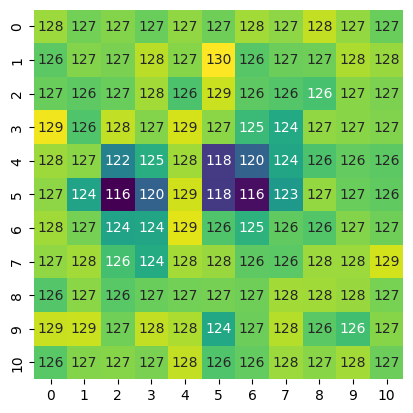} \\
    \end{tabular}
    \caption{
        Results of training from a combined grid points with significant leakage on the software AES128 evaluated on one of the datasets while trained on the other.
        Values are mean ranks of 5,000 holdout traces targeting the index 11 of the S-BOX input in the first round.
        The captions are which dataset has been used to evaluate.
    }
    \label{tab:sw_aes_multiplace_both}
\end{table}

\Cref{tab:sw_aes_multiplace_both} clearly shows there is leakage present even in the portability scenario of profiling on one chip and evaluating on another.
However we do see a significant performance drop compared to \Cref{tab:sw_aes_grid_leakages}.

\subsection{Amount of Training Data}\label{sec:amount_of_training_data}

The usual metric in SCA is the number of traces needed to reveal some non-trivial information.
In the particular case of profiling attacks we usually talk about a tradeoff between the amount of training data and attack/evaluation data.
In either case one attempts to use as few data as possible.
Deep learning on the other hand scales with additional good quality data.
When training with all the data from multiple $(x,y,z)$ coordinates we inadvertently use more data during the profiling phase.
To fairly compensate for this fact we redo some of our experiments with the same or larger amount of training data.
This helps us to distinguish whether we are benefiting more from the amount of training data or more from the diversity of the data.
Note that we have already captured more data and, in practice, we do not want to waste that effort.

To tackle this potential issue we captured a large set of training traces at the closer position of the \datasetSwAesZrh{} hot-spot (position $(6,5)$ in \Cref{tab:sw_aes_grid_leakages}).
Concretely we captured 7.9M training traces which corresponds to the amount of traces captured in the closer half of \datasetSwAesZrh{}, e.g., a half of the total size.
Thus roughly five times more data than multi-place training from the closer half of \datasetSwAesZrh{} (and $3.5$-times more data than multi-place training combining close and far EM probe positions).
Due to the amount of data we were also able to train much more (50,000 epochs vs 150 epochs) while still improving the validation accuracy.
The result of cross-device evaluation is shown in the first row of \Cref{tab:sw_aes_hotspot_vs_multiplace}.
In the portability setting the multi-place training clearly wins over the hot-spot which can produce significant leakage in just one of the grid points.

\begin{table}
    \centering
    \setlength{\tabcolsep}{0pt}  % No spacing between columns
    \def\figWidth{.24\linewidth}  % width of each figure
    \def\rotateOrigin{t}
    \begin{tabular}{cccc}
         Multi-place/close &
         Multi-place/far &
         Hot-spot/close &
         Hot-spot/far \\
    \multicolumn{4}{c}{Evaluated on \datasetSwAesMun{}:} \\
         \includegraphics[width=\figWidth]{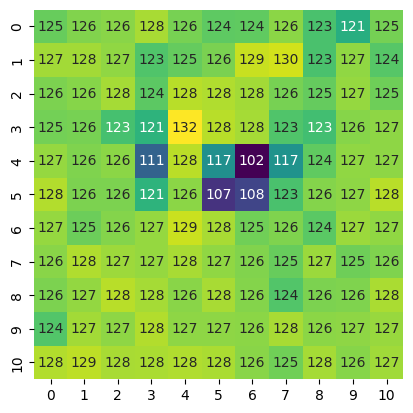} &
         \includegraphics[width=\figWidth]{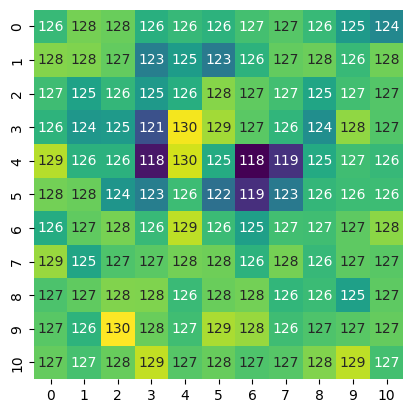} &
         \includegraphics[width=\figWidth]{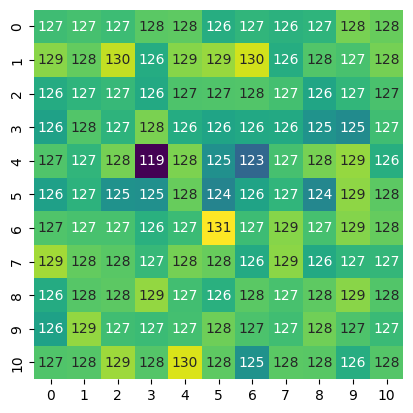} &
         \includegraphics[width=\figWidth]{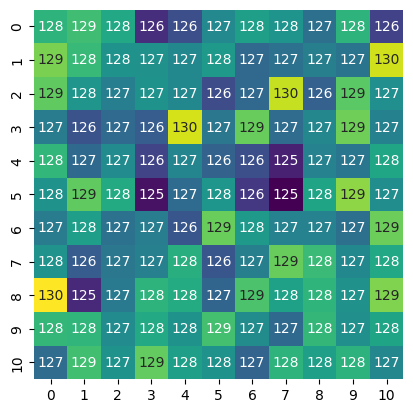} \\
    \multicolumn{4}{c}{Evaluated on \datasetSwAesZrh{}:} \\
         \includegraphics[width=\figWidth]{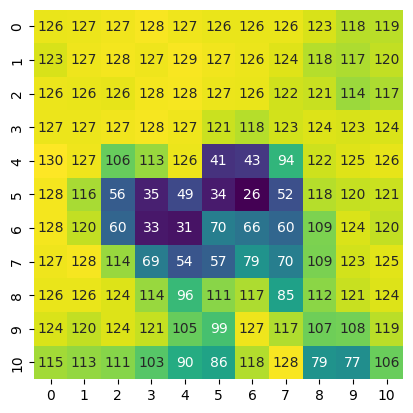} &
         \includegraphics[width=\figWidth]{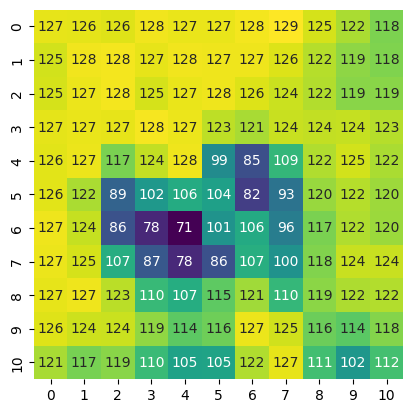} &
         \includegraphics[width=\figWidth]{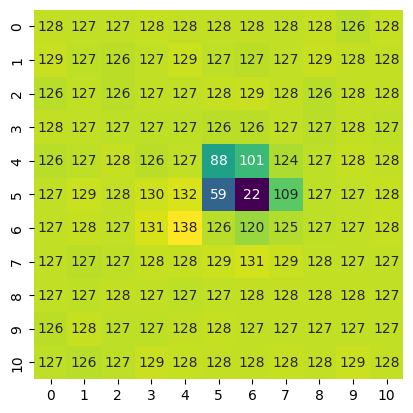} &
         \includegraphics[width=\figWidth]{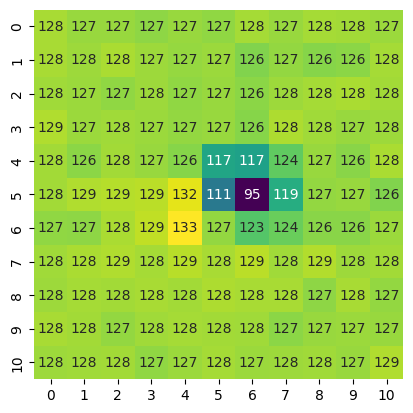} \\
    \end{tabular}
    \caption{
        Comparison of multi-place training (only the closer leakage positions) against training from the hot-spot while using much more data.
        The description is training method, either multi-place or hot-spot / which distance has been evaluated, either close or far.
        All training has been done on \datasetSwAesZrh{} or the single place hot-spot version.
    }
    \label{tab:sw_aes_hotspot_vs_multiplace}
\end{table}

Another aspect of comparing the training with the comparable amount of data is how much does the data diversity contribute to better results during displacement.
The second row of \Cref{tab:sw_aes_hotspot_vs_multiplace} shows that training on a hot-spot is significantly more accurate than multi-place trained network on the hot-spot position (no displacement at all).
But that is the only point where hot-spot training is better.
The hot-spot trained model is much more sensitive to the~$x,y$ displacement.
Additionally the hot-spot trained model is also very sensitive to~$z$ displacement compared to the multi-place model which was also trained only traces from the closer distance to the chip.
Comparing the multi-place training purely to \Cref{tab:sw_aes_grid_leakages} which had much fewer examples to train with would be unfair.

\subsection{Interpolation}\label{sec:interpolation}

Our claims of position resilience might not be sound in the extremely unlikely case that two grids of two different setups would overlap exactly or almost exactly.
In such a case we would have shown only device to device transfer on the grid points but perhaps not inbetween these grid points.
To avoid capturing an excessive amount of data we capture just 1,024 traces from each fine grid point and plot heatmap of average rank.
\Cref{tab:sw_aes_very_fine} shows smooth leakage values between the original training grid points.

It is clear that the multi-place trained cross-device evaluated neural network is able to detect leakage where the hot-spot trained model does not.
The leakage is much smaller due to using less data (discussed in \Cref{sec:amount_of_training_data}) and a different measure setup.

\begin{table}
    \centering
    \def\figWidth{.3\linewidth}  % width of each figure
    \def\rotateOrigin{t}
    \begin{tabular}{ccc}
         Multi-place (close) \datasetSwAesZrh{} &
         Multi-place (close) \datasetSwAesMun{} &
         Hot-spot training \\
         % There are also heatmaps with automatic vmin vmax (which differs between evaluations). But this is rather misleading.
         \includegraphics[width=\figWidth]{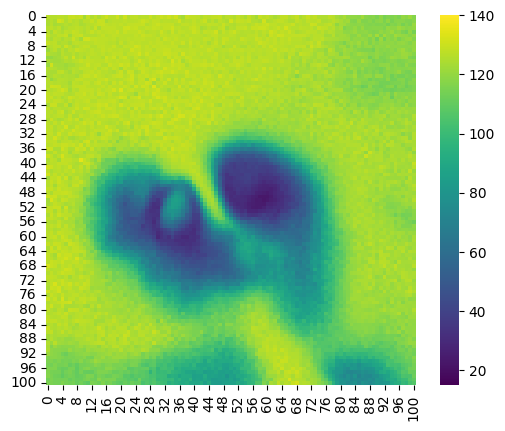} &
         \includegraphics[width=\figWidth]{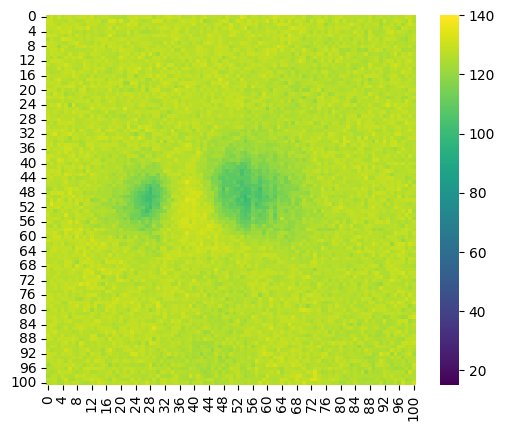} &
         \includegraphics[width=\figWidth]{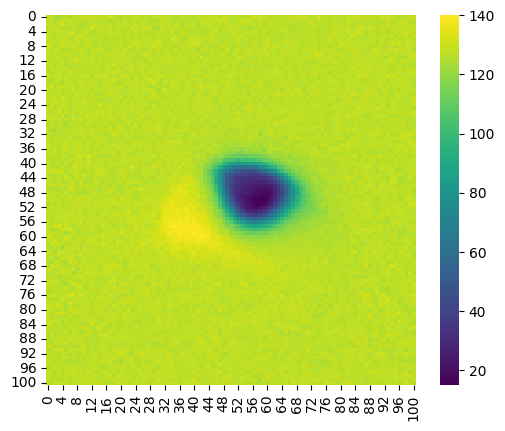} \\
         % Show threshold of 120
         \includegraphics[width=\figWidth]{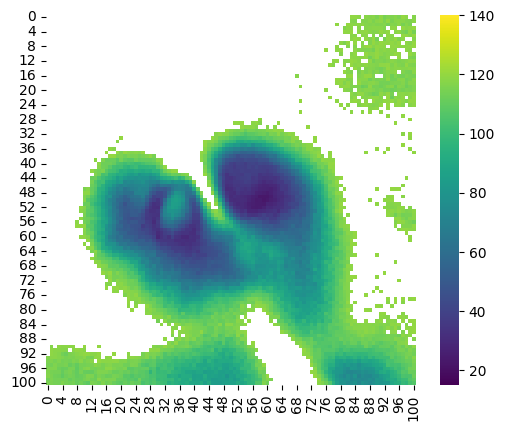} &
         \includegraphics[width=\figWidth]{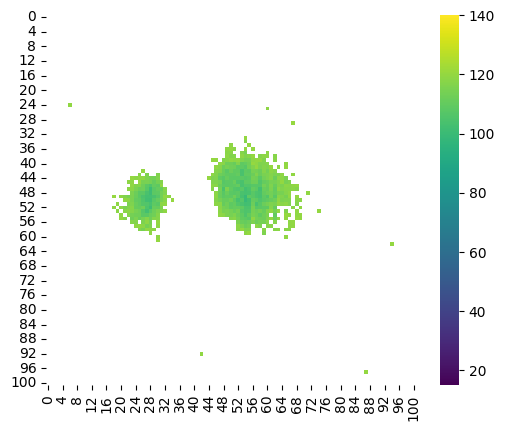} &
         \includegraphics[width=\figWidth]{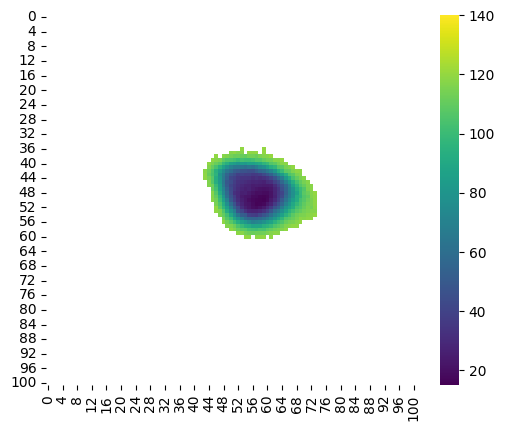} \\
    \end{tabular}
    \caption{
        A fine grid of $101 \times 101$ grid points closer to the chip covering an area of $5mm \times 5mm$ with a step of $0.05mm$.
        Evaluations of mean rank comparing multi-place training with training from a hot-spot.
        Note that the middle column corresponds to profiling on \datasetSwAesMun{} data while evaluating on \datasetSwAesZrh{}.
        The other two columns correspond to both profiling and evaluating on \datasetSwAesZrh{}.
        The last row shows where the mean rank is below the threshold 120 (fully random is 127.5).
    }
    \label{tab:sw_aes_very_fine}
\end{table}

\subsection{Distance from the Device Under Test}\label{sec:distance_from_dut}

The inverse-square law tells us that the intensity of EM signal is proportional to the reciprocal of the square of the distance.
Thus the farther the EM probe the smaller signal we expect.
We show that our analysis is independent of the precise probe distance to the extent permitted by the inverse-square law.
To avoid the possibility that the two labs by chance managed to place the probe in the exact same distance we use the closer SW-AES datasets to train and evaluate on the farther and vice versa.
Using a 3D-positioning table placing the probe tip within $0.25mm$ of the chip is possible using a naked eye.
\Cref{tab:sw_aes_multiplace_distance_vs_other_distance} shows that when training multi-place with constant distance then change of device and EM probe distance of at least $0.25mm$ does not completely distort the signal.

\begin{table}
    \centering
    \setlength{\tabcolsep}{0pt}  % No spacing between columns
    \def\figWidth{.24\linewidth}  % width of each figure
    \def\rotateOrigin{t}
    \begin{tabular}{cccc}
         \datasetSwAesMun{} C/F &
         \datasetSwAesMun{} F/C &
         \datasetSwAesZrh{} C/F &
         \datasetSwAesZrh{} F/C \\
         \includegraphics[width=\figWidth]{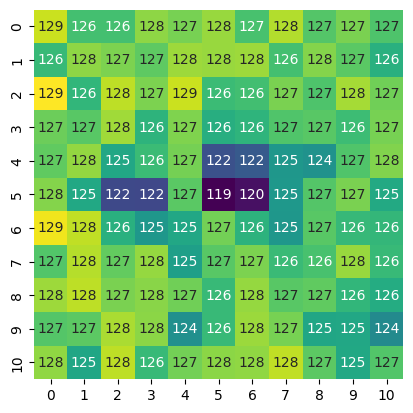} &
         \includegraphics[width=\figWidth]{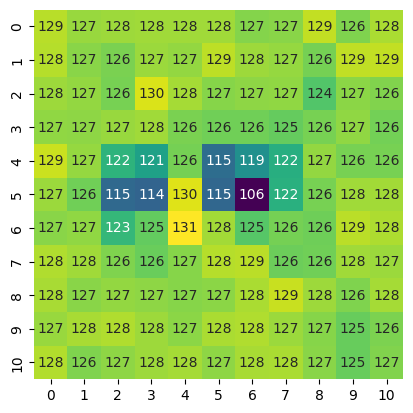} &
         \includegraphics[width=\figWidth]{figs/sw_multiplace/evaluations/heatmap_res_pp_paper_swaes_zrh_close_1748015432.keras_mun_3d_grid_full_far_mean_rank.png} &
         \includegraphics[width=\figWidth]{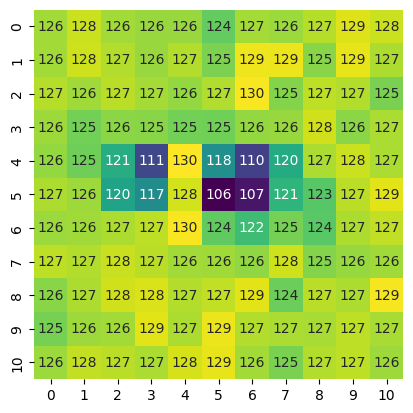} \\
    \end{tabular}
    \caption{
        Mean rank when training multi-place on one distance of either \datasetSwAesMun{} or \datasetSwAesZrh{} and evaluating at another distance of the other dataset.
        In both datasets the close and far are at distance $0.25mm$.
        Thus at least one of the heatmaps corresponds to training on one device and evaluating on another where the EM probe distance from the device changes by at least $0.25mm$.
        To allow concise representation we use the following notation: dataset $D1/D2$ meaning we train multi-place on the given dataset with distance $D1$ and evaluate on the other dataset at distance $D2$ where C means close and F far.
    }
    \label{tab:sw_aes_multiplace_distance_vs_other_distance}
\end{table}

\subsection{Training with All Data}\label{sec:training_all_data}

Given the result of this section one might also ask what happens when we would train on all captured data.
Figure~\ref{tab:sw_aes_all_data} shows that using a lot of data with very small leakage information does not bring a great advantage.
We can see that the top leakage is lower compared to selecting grid points with significant leakage to train.
On the other hand we see slightly larger resilience to the position as a small but noticeable leakage is being detected even along the grid edges.

\begin{table}
    \centering
    \setlength{\tabcolsep}{0pt}  % No spacing between columns
    \def\figWidth{.48\linewidth}  % width of each figure
    \def\rotateOrigin{t}
    \begin{tabular}{cc}
         Trained \datasetSwAesMun{}{} &
         Trained \datasetSwAesZrh{}{} \\
         \includegraphics[width=\figWidth]{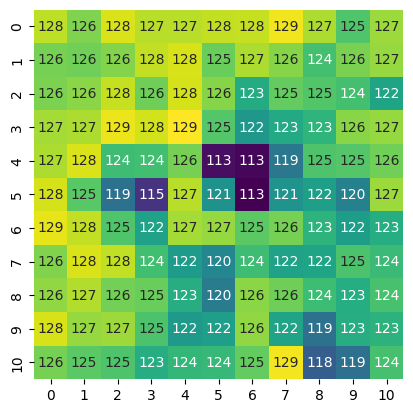} &
         \includegraphics[width=\figWidth]{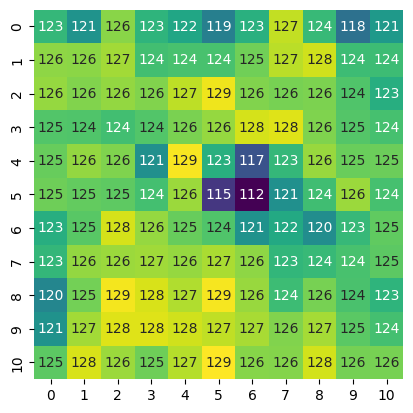} \\
    \end{tabular}
    \caption{
        Training with all data on one dataset and evaluating on the other dataset.
        Compare with \Cref{tab:sw_aes_multiplace_both} where just grid points with significant leakage were used.
    }
    \label{tab:sw_aes_all_data}
\end{table}

\subsection{Hardware Accelerated AES Hybrid Approach}\label{sec:hwaes_hybrid_approach}

One may face different challenges when considering hardware accelerated implementations.
We thus repeat some of our experiments with STM32F4 hardware accelerated AES.
As suggested by \cite{cwTutorialSCA201Lab22} we target the last round Hamming distance.
Due to the low leakage and thus poor deep learning metrics we first moved from classification of last round Hamming distance to a regression formulation.
The neural network instability during classification manifested itself by predicting either zero or eight as the Hamming distance in majority of predictions and even these two classes would be greatly biased.
Reinterpreting the problem as regression was not enough and we saw just minute improvements over random results.
For our evaluation we resorted to a hybrid approach of using the neural network as a signal amplifier.
We feed a trace to the neural network which outputs 16 floating point numbers representing the 16 Hamming distances in the last round (that is 16 times a regression instead of classification).
The 16 floats would then be fed into a classical CPA instead of the original trace during the attack phase.
This hybrid approach works remarkably well in our setting -- one can compare results of \Cref{tab:hw_aes_train_grid,tab:hwAesHoldoutGrid} with \Cref{tab:cpaHwAes}.
To the best of our knowledge this is the first use of this technique.
The closest to this technique might be the paper \cite{lee2025leakdit} where the authors use a denoising diffusion transformer followed by a classification convolutional neural network.
In our case the classification part is CPA.

We reduce the number of our experiments and do not repeat all of them for hardware accelerated AES.
The grids in \Cref{tab:hw_aes_train_grid} show that at least on the training chip there is detectable leakage everywhere.

\begin{table}
    \centering
    \setlength{\tabcolsep}{0pt}  % No spacing between columns
    \def\figWidth{.48\linewidth}  % width of each figure
    \def\rotateOrigin{t}
    \begin{tabular}{cc}
         Traces until disclosure [thousands] &
         Average rank after 128k traces \\
         \includegraphics[width=\figWidth]{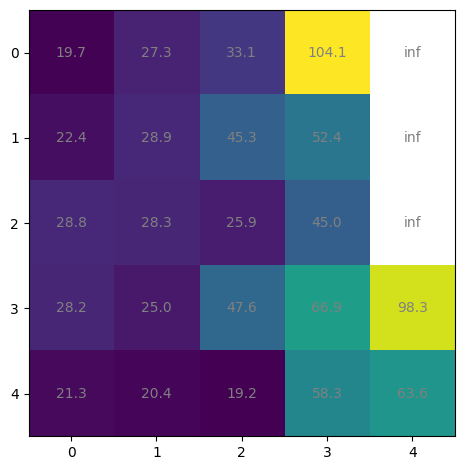} &
         \includegraphics[width=\figWidth]{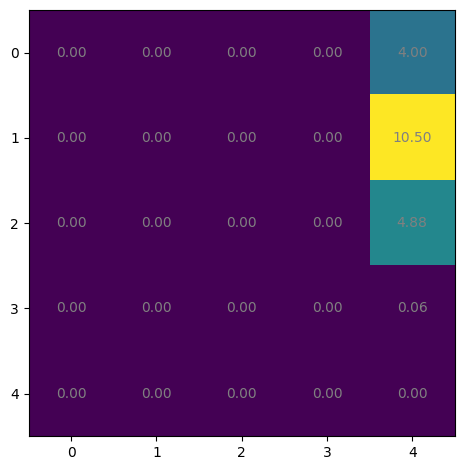} \\
    \end{tabular}
    \caption{
        Evaluation of the hybrid approach attack trained on \datasetHwAesTrain{} evaluated on the holdout part of the same dataset.
        The left grid shows how many thousand traces were needed to fully disclose the secret key in case the key was not fully recovered in 128k traces we write infinity.
        We also provide the average rank of the correct byte value after 128k traces to identify if there was at least partial leakage.
    }
    \label{tab:hw_aes_train_grid}
\end{table}

On the \datasetHwAesHoldout{} dataset we have not achieved full key discovery in 128k traces.
Thus we recaptured a similar \datasetHwAesHoldoutLarger{} with significantly more traces per position and a larger area.
The probe position between \datasetHwAesHoldout{} and \datasetHwAesHoldoutLarger{} has been moved.
However we have not retrained the model again (it was trained on data from a different chip and room anyway).

\begin{table}
    \centering
    \setlength{\tabcolsep}{0pt}  % No spacing between columns
    \def\figWidth{.48\linewidth}  % width of each figure
    \def\rotateOrigin{t}
    \begin{tabular}{cc}
         Traces until disclosure [thousands] &
         Average rank after 742k traces \\
         \includegraphics[width=\figWidth]{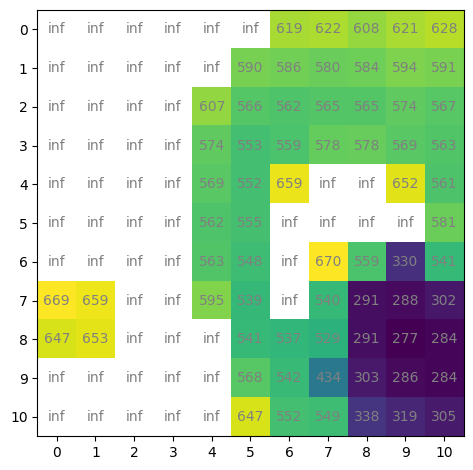} &
         \includegraphics[width=\figWidth]{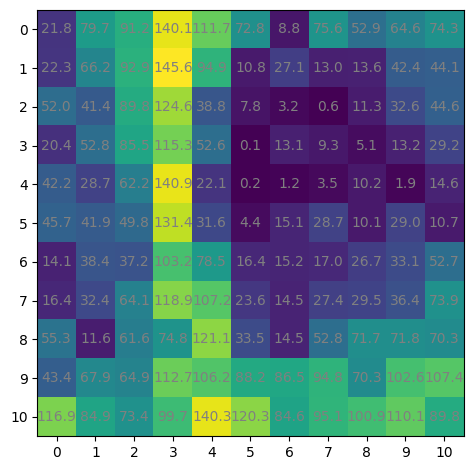} \\
    \end{tabular}
    \caption{
        Evaluation of the hybrid approach attack trained on \datasetHwAesTrain{} evaluated on the \datasetHwAesHoldoutLarger{}.
        The grid step is $0.25mm$ covering the total grid area of $2.5 \times 2.5 mm^2$.
    }
    \label{tab:hwAesHoldoutGrid}
\end{table}

\subsection{Multi-place Training with the \cite{ninan2024second} Datasets}\label{sec:dl_results_second_look}

Evaluating multi-place training on \datasetSecondLookXMEGA{} and \datasetSecondLookSTM{} can be done only by comparing how much leakage we detect on the holdout part which is a single grid point.
We thus opt to revert the meaning of dataset splits of~\cite{ninan2024second}.
We train on their \datasetSecondLookXMEGA{} X2 and \datasetSecondLookSTM{} S2 which are $3 \times 3$ grids (135k training, 15k validation).
And reserve the single grid point from the X1 and S1 chips for holdout (150k traces).
Unfortunately this makes our results not directly comparable since our training and validation splits are swapped.
But overall whenever \cite{ninan2024second} saw leakage on a grid point we were able to train on that grid point and see leakage on the other chip.

\begin{table}
    \centering
    \setlength{\tabcolsep}{0pt}  % No spacing between columns
    \def\figWidth{.24\linewidth}  % width of each figure
    \def\rotateOrigin{t}
    \begin{tabular}{cccc}
         \datasetSecondLookXMEGA{} X2 &
         \datasetSecondLookXMEGA{} X1 &
         \datasetSecondLookSTM{} S2 &
         \datasetSecondLookSTM{} S1 \\
         \includegraphics[width=\figWidth]{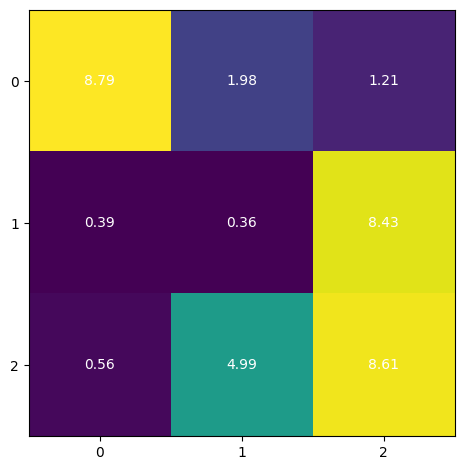} &
         \includegraphics[width=\figWidth]{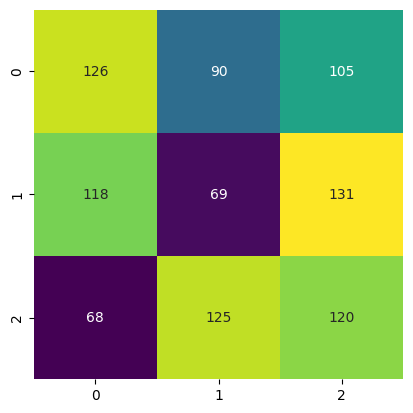} &
         \includegraphics[width=\figWidth]{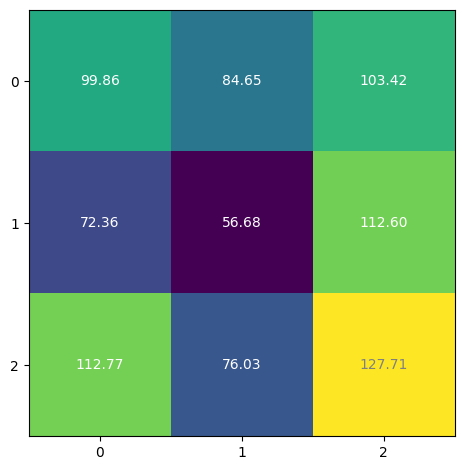} &
         \includegraphics[width=\figWidth]{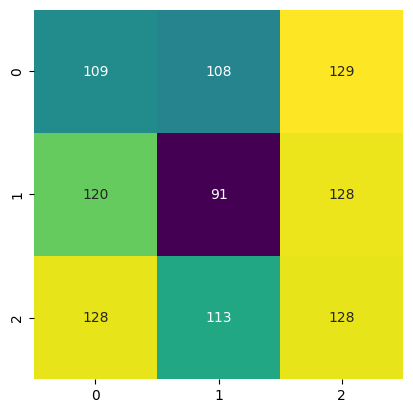} \\
    \end{tabular}
    \caption{
        Mean rank of S-Box input byte 0 on the \datasetSecondLookXMEGA{} and \datasetSecondLookSTM{} datasets of \cite{ninan2024second}.
        We use the X2 and S2 data for training and the X1 and S1 data for holdout.
    }
    \label{tab:second_look}
\end{table}

\Cref{tab:second_look} shows single grid point leakages in validation and holdout data.
Interestingly we see a lot of accuracy decrease in the cross-chip detected leakage levels in \datasetSecondLookXMEGA{} but less in the \datasetSecondLookSTM{}.
It is likely that our deep learning model specialized too much on the details of the training chip traces.

\begin{table}[]
    \centering
    \begin{tabular}{lllllllll}
\hline
 Combined positions   & 2             & 3             & 4             & 5             & 6             & 7             & 8             & 9             \\
\hline
 Limited data XMEGA   & \textbf{67.4} & 70.5          & \textbf{17.8} & \textbf{37.9} & \textbf{33.2} & \textbf{40.3} & \textbf{28.0} & \textbf{30.0} \\
 All data XMEGA       & \textbf{39.3} & \textbf{50.8} & \textbf{28.7} & \textbf{53.2} & \textbf{43.2} & \textbf{34.4} & \textbf{38.2} & \textbf{38.9} \\
 Limited data STM     & \textbf{89.4} & \textbf{78.7} & \textbf{83.8} & \textbf{80.2} & 127.4         & 127.7         & \textbf{74.1} & \textbf{73.1} \\
 All data STM         & \textbf{77.8} & \textbf{71.2} & \textbf{71.5} & \textbf{74.1} & \textbf{70.5} & \textbf{71.5} & \textbf{75.3} & \textbf{72.7} \\
\hline
\end{tabular}
    \caption{
      The mean rank on the holdout chip (either X1 or S1) when training on data from multiple places of the training chip (either X2 or S2).
      In each experiment we combine $n$ positions with the best performance on the validation split (either X2 or S2, see \Cref{tab:second_look}).
      We either use all available data for training or limit the data to at most 135k examples similarly to the one position training.
      Better results than single point training (either lower than 68 for XMEGA and 91 for STM) are shown in bold.
    }
    \label{tab:second_look_n_best}
\end{table}

\Cref{tab:second_look_n_best} shows what happens when we use traces from the top $n$~grid positions.
We either use all data or compensate by taking at most 135k traces similar to single grid point training.
In almost all cases multi-place training outperforms single grid point training.
In the case of \datasetSecondLookSTM{} using all data consistently gives better results up to one case where the result is close.
This is consistent with the expectation of deep learning scaling with more data.
Surprisingly in the case of \datasetSecondLookXMEGA{} we see that limiting the amount of data gives a better result in more than half cases.
We have no good explanation for this phenomenon.
We just remark that limiting or not limiting the amount of data does not seem to affect the validation performance.

\section{Classical Attacks}\label{sec:classical_attacks}

Prior research has evaluated classical attacks in the setting of portability and position resilience as discussed in \Cref{sec:related_work}.
However to the best of our knowledge no experiments with multi-place profiling have been done.
Classical profiled attacks such as template attacks~\cite{CHES:ChaRaoRoh02} greatly benefit from trace alignment and ideally picking of points of interest.
But they would also suffer from differences caused by different EM probe position.
Our traces are synchronized by the trigger and one could in principle align them further using known techniques.
In this section we provide results of classical techniques to compare observed leakage.
Thanks to the unsupervised nature of CPA~\cite{CHES:BriClaOli04} we can estimate how much leakage there is in each of the grid points.

We present heatmaps of the average rank of the correct key byte value after the whole holdout set and or the number of traces until full key disclosure whichever makes more sense for the concrete dataset.
We avoid using the guessing entropy defined by~\cite{EC:StaMalYun09} since we would be either providing estimates or need to exactly specify the enumeration procedure for key guesses.
We rather mark the number of traces until complete key disclosure or provide the average rank of the correct key byte value.
A realistic attacker could choose to trade-off the amount of data for more compute.

We omit classical experiments on the \Cref{tab:second_look} since those are well studied already in the original publication.

\subsection{SW AES}\label{sec:classicalAttacksSwAes}

\Cref{tab:sw_aes_cpa} is showing a much larger area where leakage is detectable on the \datasetSwAesZrh{} compared to the \datasetSwAesMun{}.
This is due to the \datasetSwAesZrh{} having more than four times as many traces per grid point than \datasetSwAesMun{}.
This was caused by misspecification of the amounts of traces.
The capture setups were altered before we noticed this difference.
We decided against artificially limiting the amount of traces as more traces give us a better idea about the present leakage.
Another difference is that \datasetSwAesMun{} holdout is not fixed key and thus we needed to simulate a fixed key by changing the plaintext for the use of~CPA.

\begin{table}
    \centering
    \setlength{\tabcolsep}{0pt}  % No spacing between columns
    \def\figWidth{.24\linewidth}  % width of each figure
    \def\rotateOrigin{t}
    \begin{tabular}{cccc}
    \datasetSwAesMun{} close & \datasetSwAesMun{} far & \datasetSwAesZrh{} close & \datasetSwAesZrh{} far \\
    \multicolumn{4}{c}{Thousands of holdout traces until full key disclosure:} \\
    \multicolumn{2}{c}{inf is more than $6,000$} & \multicolumn{2}{c}{inf is more than $25,600$} \\
         \includegraphics[width=\figWidth]{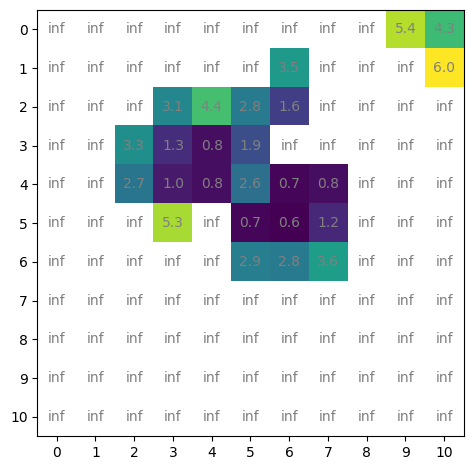} &
         \includegraphics[width=\figWidth]{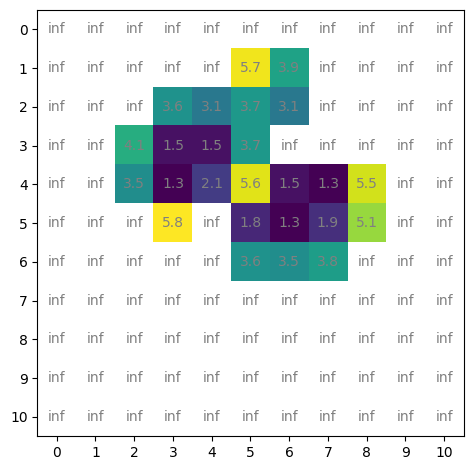} &
         \includegraphics[width=\figWidth]{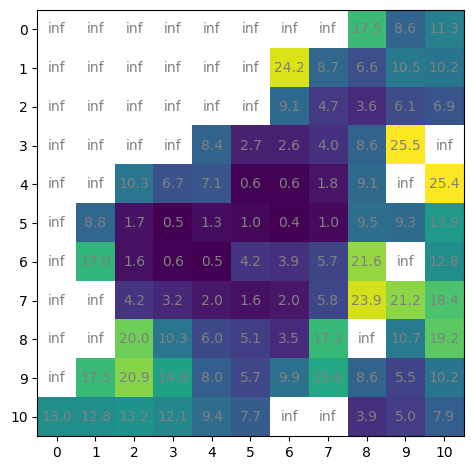} &
         \includegraphics[width=\figWidth]{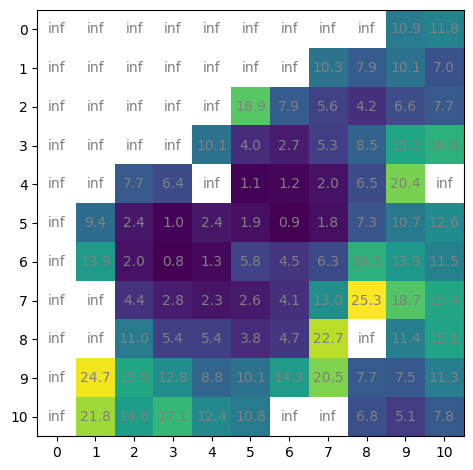} \\
    \multicolumn{4}{c}{Average rank of the correct byte value after} \\
    \multicolumn{2}{c}{$6,000$ holdout traces:} & \multicolumn{2}{c}{$25,600$ holdout traces:} \\
         \includegraphics[width=\figWidth]{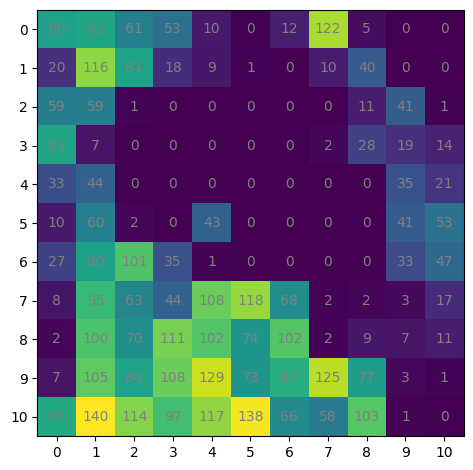} &
         \includegraphics[width=\figWidth]{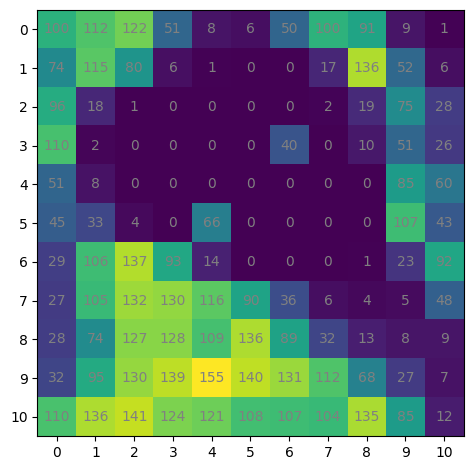} &
         \includegraphics[width=\figWidth]{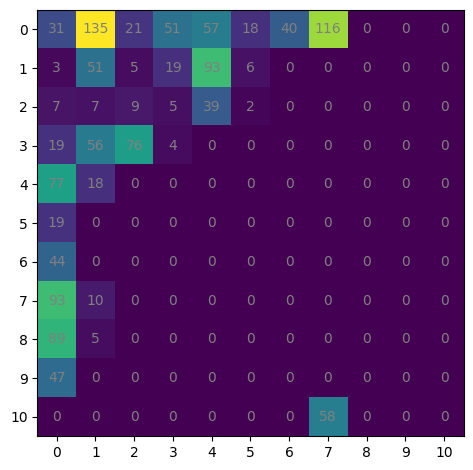} &
         \includegraphics[width=\figWidth]{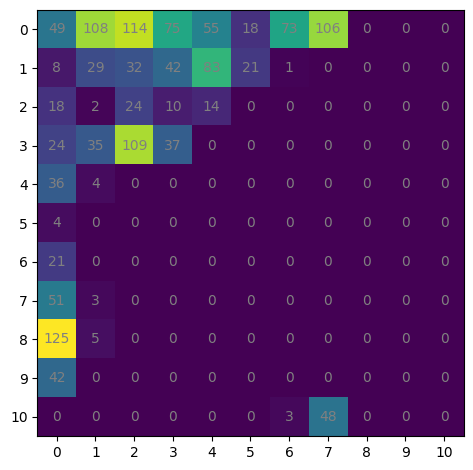} \\
    \end{tabular}
    \caption{
        The number of traces needed until full key disclosure (in thousands).
        Whenever the full key was not fully recovered using the amount of holdout traces we note it as \texttt{inf}.
        The average key byte rank is shown in the second row.
    }
    \label{tab:sw_aes_cpa}
\end{table}

\subsection{HW AES}\label{sec:classicalAttacksHwAes}

\Cref{tab:cpaHwAes} shows that CPA was rather inefficient against the HW accelerated AES implementation.

\begin{table}
    \centering
    \setlength{\tabcolsep}{0pt}  % No spacing between columns
    \def\figWidth{.31\linewidth}  % width of each figure
    \def\rotateOrigin{t}
    \begin{tabular}{cc}
         \datasetHwAesHoldout{} (128k traces) &
         \datasetHwAesHoldoutLarger{} (742k traces) \\
         \includegraphics[width=0.31\linewidth]{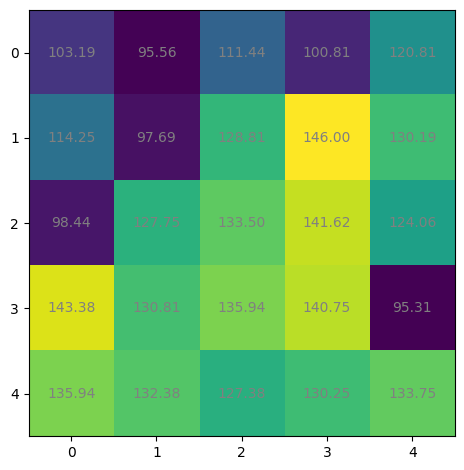} &
         \includegraphics[width=0.48\linewidth]{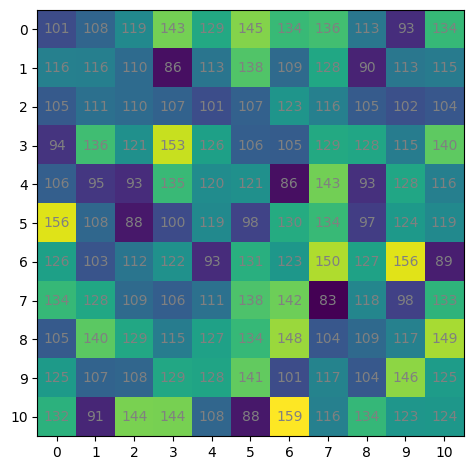}
    \end{tabular}
    \caption{
        CPA on hardware accelerated AES where we provide the average rank of the correct byte value after given number of traces (in thousands) to identify if there was at least partial leakage.
        We target the Hamming weight of the last round diff similarly to our ML experiment.
    }
    \label{tab:cpaHwAes}
\end{table}
\section{Conclusion and Future Directions}\label{sec:conclusion}

Repeatability and transferability of attacks from one device to another has not been in an ideal state especially for the electromagnetic emanations.
We provide a new set of large scale datasets captured by different teams and in different places to extend on pioneering work of portability of profiled side channel analysis.

\Cref{sec:dl_results} shows that training on EM traces captured from multiple different positions gives more position resilient results when evaluated on traces from another chip.
This is very promising since it provides better results when evaluating on a different chip while using roughly the same amount of profiling data.
Moreover we are no longer forced to throw away data captured during the initial hot-spot search.
These experiments confirm that repositioning over the attack chip is possible using just a naked eye and a photo of the other setup.

Comparing the promising interpolation and extrapolation results of \Cref{sec:interpolation} and comparable but slightly lower maximal leakage when training will all data (\Cref{sec:training_all_data}).
One could suggest more elaborate schemata for selecting training data.
\Cref{sec:classicalAttacksSwAes} shows that there is leakage in a larger area than we used for our multi-place training.
Figuring out how to use traces from grid points with lower level of signal might provide much more information and even more resilience during repositioning.
One could first train on places with more signal and slowly mix in traces from grid points with lower signal similarly to~\cite{DBLP:conf/iclr/LoshchilovH17}.
Very fine grids for training data acquisition would provide smoother interpolation of position-dependent trace changes.
This might have positive effect on both training and generalizability.
We leave these experiments as an interesting future direction.

Due to our positioning equipment we have trained from regularly spaced grid points.
This provided us with great deal of precision when evaluating the attack phase and doing ablation studies since we knew exactly where the holdout traces have been captured.
But not all labs have precise positioning devices due to their high price or due to another experiment using them.
Our multi-place training was benefiting already when traces were coming from ten or twenty grid positions.
The interpolation experiments of \Cref{sec:interpolation} suggest that there is nothing special about grid points in the attack phase.
It is thus plausible to assume that one would get similar results using a hand positioned probe holding device and repositioning it slightly several times during capture.
An experiment is needed to confirm this intuition.

Due to the amount of captured data we have focused only on variants of AES.
However extending these experiments to other cryptographic primitives would be an interesting research in itself.

%%%% 8. BILBIOGRAPHY %%%%
\bibliographystyle{alpha}
\bibliography{abbrev3,crypto,references}

%\newpage \listoftodos

\end{document}